\documentclass[prl,showpacs,twocolumn,floatfix]{revtex4}

\usepackage{graphicx}
\usepackage{latexsym}
\usepackage{eso-pic}
\usepackage{type1cm}
\usepackage{color}
\usepackage{amsmath}

\begin{document}

\title{Non-equilibrium relaxation and critical aging for driven Ising lattice 
       gases}

\author{George L. Daquila} \email{gdaquila@vt.edu}
\author{Uwe C. T\"auber} \email{tauber@vt.edu}
\affiliation{Department of Physics, Virginia Tech, 
             Blacksburg, Virginia 24061-0435}


\begin{abstract}
We employ Monte Carlo simulations to study the non-equilibrium relaxation of 
driven Ising lattice gases in two dimensions.  
Whereas the temporal scaling of the density auto-correlation function in the
non-equilibrium steady state does not allow a precise measurement of the 
critical exponents, these can be accurately determined from the aging scaling 
of the two-time auto-correlations and the order parameter evolution following a 
quench to the critical point.  
We obtain excellent agreement with renormalization group predictions based on
the standard Langevin representation of driven Ising lattice gases.
\end{abstract}


\date{\today}

\maketitle

Driven diffusive systems represent paradigmatic models that display non-trivial
non-equilibrium stationary states, and hence constitute crucial test cases for 
various analytical and numerical approaches \cite{beatebook,marrobook}.  
Already driven lattice gases with mere site exclusion and periodic boundary 
conditions yield anisotropically scale-invariant steady states.  
In one dimension, the asymptotic scaling properties of these asymmetric 
exclusion processes are governed by fluctuation-controlled exponents that 
differ from mean-field predictions; they belong to the noisy Burgers or 
one-dimensional Kardar--Parisi--Zhang (KPZ) universality class 
\cite{Forster,kpz}.  
Related models have found varied applications in the theoretical description of
biological systems \cite{chou2011}.  
Adding nearest-neighbor attractive Ising interactions, as proposed by Katz, 
Lebowitz, and Spohn (KLS) \cite{katz1983,katz1984} to describe ionic 
conductors, promotes clustering of empty and filled sites, inducing a 
continuous non-equilibrium phase transition between a disordered and a striped 
phase in $d \geq 2$ dimensions.  
In the absence of the drive, one recovers the equilibrium Ising model 
ferromagnetic phase transition.
Yet the correct long-wavelength description of the KLS Ising driven lattice gas
at the critical point has been subject to a lasting debate \cite{leung1991,wang1996,schmittmann2000,achahbar2001,caracciolo2003,caracciolo2004_a}.
Ambiguities in determining the critical exponents are in part caused by 
exceedingly slow crossovers to the asymptotic regime \cite{caracciolo2004_a}; 
also, it is crucial to implement the proper anisotropic lattice scaling, and 
infer scaling exponents 
sufficiently close to the critical point \cite{leung1996}.

Intriguing universal scaling features moreover arise in the non-equilibrium 
relaxation regime from an initial configuration that drastically differs from 
the final stationary state \cite{pleimbook}.  
When the order parameter field is conserved, one may relate the ensuing aging 
scaling exponents to those describing the non-equilibrium steady state 
\cite{janssen1989}.  
Investigating the aging and initial-slip scaling regimes thus provides an 
independent and often more easily accessible means to determine the asymptotic 
critical exponents \cite{zheng1998}.
We recently utilized the aging relaxation scaling to accurately confirm the 
exponent values for asymmetric exclusion processes \cite{daquila2011}; for a 
related study of the KPZ equation, see Ref.~\cite{henkel2011}. 
Critical quenches and the ensuing aging scaling were also analyzed for the
continuous phase transition in the contact process 
\cite{baumann2005,baumann2007}.  
Here, we employ Monte Carlo simulations to study the non-equilibrium relaxation 
properties of the KLS model following a sudden quench from the high-temperature 
disordered phase to the critical point.  
We utilize the two-time density auto-correlation function and the temporal 
evolution of the order parameter in the aging regime of large systems to 
cleanly measure the associated critical exponents in two dimensions, see also
Refs.~\cite{albano2002,albano2011}.  

The KLS model consists of $N$ hard-core particles on a 
$L_\parallel \times L_\perp^{d-1}$ lattice with periodic boundary conditions;
``$\parallel$'' and ``$\perp$'' denote directions parallel and perpendicular to
the drive. 
The occupation number $n_{\bf i}$ at any lattice site ${\bf i}$ is restricted to
$0$ or $1$; multiple site occupations are prohibited.
In addition to this on-site exclusion, we consider an attractive 
nearest-neighbor Ising interaction 
$H_{\rm int} = - J \sum_{\langle {\bf i,j} \rangle} n_{\bf i} \, n_{\bf j}$, $J > 0$.
The transition rate evolving the system from configuration $X$ to $Y$ then is
\begin{equation}
\label{hopping_rates}
  R(X \to Y) \propto e^{- \beta \left[ H_{\rm int}(Y) - H_{\rm int}(X) + \ell E \right]}
\end{equation}
with inverse temperature $\beta$; $\ell = \{ -1,0,1 \}$ indicates particle hops
against, transverse to, and along with the drive.
The bias $E$ and periodic boundary conditions generate a non-zero particle 
current through the system; since the rates (\ref{hopping_rates}) do not 
satisfy detailed balance, the system will eventually settle in a 
non-equilibrium steady state.
For $E = 0$, and with $n_{\bf i} = \frac12 (\sigma_{\bf i} + 1)$, the KLS lattice
gas reduces to the equilibrium Ising model with spin variables 
$\sigma_{\bf i} = \pm 1$ and conserved Kawasaki kinetics.
We employ the standard Metropolis algorithm with rates (\ref{hopping_rates}),
setting $J = 1$ and limiting our study to the case $E \to \infty$, implying 
that particle jumps with $\ell = -1$ are strictly forbidden.
Time is measured in Monte Carlo sweeps, wherein on average one update per
particle is attempted.

The theoretical treatment of the long-time and large-scale properties of the 
KLS model uses a continuum description in terms of a density field 
$\rho(\vec{x},t)$ or equivalently the conserved spin density 
$\phi(\vec{x},t) = 2 \rho(\vec{x},t) - 1$. 
Janssen and Schmittmann \cite{janssen1986} as well as Leung and Cardy 
\cite{leung1986} constructed a mesoscopic Langevin equation for the critical 
KLS model near the critical point,
\begin{align}
\label{lang_crit}
  &\partial_t \phi(\vec{x},t) = \frac{\mathcal{E}}{2} \, \nabla_\parallel 
  \phi(\vec{x},t)^2 - \vec{\nabla}_\perp \cdot \vec{\xi}_{\perp}(\vec{x},t) \\ 
  &+ \lambda \left[ \left( c \, \nabla^2_\parallel + \tau \, 
  \vec{\nabla}^2_\perp - \vec{\nabla}_\perp^4 \right) \phi(\vec{x},t) 
  + u \, \vec{\nabla}^2_\perp \phi(\vec{x},t)^3 \right] , \nonumber
\end{align}
and analyzed it by means of perturbative field-theoretic dynamic 
renormalization group (RG) methods near the upper critical dimension $d_c = 5$.
In the ordered phase, the system forms stripes oriented parallel to the drive, 
and correspondingly the critical control parameter is 
$\tau = (T-T_{\rm C})/{T_{\rm C}} \to 0$, whereas $c > 0$.  
$\mathcal{E}$ represents a coarse-grained driving field, with 
$\mathcal{E} \to {\rm const}$ as $E \to \infty$.  
Since only the transverse sector becomes critical, just the associated Gaussian
noise needs to be accounted for, with 
$\langle \xi_\perp(\vec{x},t) \rangle = 0$ and correlator
$\langle \xi_{\perp,i}(\vec{x},t) \, \xi_{\perp,j}(\vec{x}',t') \rangle = 
n_\perp \delta(\vec{x}-\vec{x}') \delta(t-t') \delta_{ij}$ 
($i,j = 1,\ldots,d-1$).

We are primarily interested in the density-density correlation function
$C(\vec{x},\vec{x}',t,s) = \langle \rho(\vec{x},t) \rho(\vec{x}',s) \rangle - 
{\bar \rho}^2$, with the mean density 
${\bar \rho} = \langle \rho(\vec{x},t) \rangle = \frac12$ in our simulations.
Assuming spatial and temporal translational invariance in the non-equilibrium 
steady state, one arrives at the general scaling form (with positive scale 
parameter $b$)
\cite{beatebook}
\begin{align}
\label{ising_full_scaling_real}
  & C(\vec{x}_\perp,x_\parallel,t,\tau,L_{\perp},L_{\parallel}) = \\ 
  &= b^{d+\Delta-2+\eta} \, C\Bigl( \vec{x}_\perp b, x_\parallel b^{1+\Delta},
  t b^z, \tau b^{-1/\nu}, L_\perp b, L_\parallel b^{1+\Delta} \Bigr) \, , 
  \nonumber
\end{align}
where $t$, $x_\parallel$, and $\vec{x}_\perp$ now represent time and position 
differences.
Here, $\nu$ and $z$ denote the transverse correlation length \cite{trcorr} and 
dynamic critical exponents, the latter associated with critical slowing-down, 
while $\eta$ characterizes the power law correlations at criticality. 
The external drive induces spatially anisotropic scaling, captured by a nonzero
anisotropy exponent $\Delta$ (in the mean-field approximation $\Delta = 1$), 
which in turn results in distinct exponents in the direction of the drive,
$\nu_\parallel = \nu (1 + \Delta)$, whence $z_\parallel = z / (1 + \Delta)$, 
and $\eta_\parallel = (\eta + 2 \Delta) / (1 + \Delta)$ \cite{beatebook}.
In the RG analysis near the upper critical dimension $d_c = 5$ as determined by 
the coupling associated with the drive $\mathcal{E}$, the static nonlinearity 
$u$ becomes (dangerously) irrelevant, and the overall conservation law together 
with Galilean invariance fix the scaling exponents to all orders in the 
$\epsilon = 5 - d$ expansion \cite{janssen1986,leung1986}. 
The resulting exact numerical values (henceforth referred to as ``JSLC'') are 
listed in Table~\ref{ising_table} for $\epsilon > 0$, and specifically for 
$d=2$.
\begin{table}[!t]
\caption{Critical exponents for the KLS and RDLG models (precise definitions
         are provided in the text).}
\label{ising_table}
\begin{center}
\begin{tabular}{|c|c|c||c|c|}
\hline
         & \multicolumn{2}{|c||}{JSLC -- exact} 
         & \multicolumn{2}{|c|}{RDLG -- $O(\varepsilon^{2})$} \\ \hline
         & $\epsilon = 5-d$ & $d=2$ & $\varepsilon = 3-d$ & $d=2$ \\ \hline
$\Delta$ & $1 + \epsilon/3$ & $2$ & $1 - 2\varepsilon^2/243$ & $0.992$\\ \hline
$z$      & $4$              & $4$ & $4 - 4\varepsilon^2/243$ & $3.984$\\ \hline
$\nu$    & $1/2$            & $1/2$ & $\frac12 + \frac{\varepsilon}{12} +
                              \frac{\varepsilon^2}{18} \bigl[ \frac{67}{108} 
              + \ln \bigl( \frac{2}{\sqrt{3}} \bigr) \bigr]$ & $0.626$\\ \hline
$\eta$   & $0$              & $0$   & $4\varepsilon^2/243$   & $0.016$\\ \hline
$\beta$  & $1/2$            & $1/2$ & $\frac{\nu}{2} \left( 1 + \Delta 
                              - \varepsilon + \eta \right)$ & $0.315$ \\ \hline
$\zeta$  & $1 - \epsilon/6$ & $1/2$ & $\bigl( 2 - \varepsilon + 
                              \frac{2 \varepsilon^2}{243} \bigr) / \bigl( 4 - 
                          \frac{4\varepsilon^2}{243} \bigr)$ & $0.253$\\ \hline
\end{tabular}
\end{center}
\end{table}

Early Monte Carlo simulation data \cite{valles1987,marro1987b,wang1989} of the 
critical KLS model with infinite drive indicated discrepancies with these 
JSLC predictions.  
Leung \cite{leung1991} subsequently applied careful anisotropic finite-size 
scaling to his Monte Carlo data, and obtained agreement with the JSLC theory,
also supported by later work \cite{caracciolo2003,caracciolo2004_a}.  
A debate on the validity and implications of his results followed 
\cite{achahbar1995,leung1996}, and it was proposed 
\cite{garrido1998,santos1999,garrido2000} that the critical KLS model at 
infinite drive might not be adequately described by Eq.~(\ref{lang_crit}) in 
the asymptotic limit, but instead by the Langevin equation for the randomly 
driven Ising lattice gas (RDLG, or two-temperature model B) 
\cite{schmittmann1991},
\begin{align}
\label{lang_crit_RDLG}
  &\partial_t \phi(\vec{x},t) = - \nabla_\parallel \xi_{\parallel}(\vec{x},t)
  - \vec{\nabla}_\perp \cdot \vec{\xi}_\perp(\vec{x},t) \\
  &+\lambda \left[ \left( c \, \nabla^2_\parallel + \tau \, 
  \vec{\nabla}^2_\perp - \vec{\nabla}_\perp^4 \right) \phi(\vec{x},t) 
  + u \, \vec{\nabla}^2_\perp \phi(\vec{x},t)^3 \right] . \nonumber
\end{align}
Here the drive term $\sim \mathcal{E}$ that is the origin of the non-vanishing 
particle current has been dropped, with the reasoning that the presence of 
anisotropic noise with different strengths $n_\perp$ and $n_\parallel$ is
supposedly more relevant in the infinite drive limit.
The fluctuations for the Langevin equation (\ref{lang_crit_RDLG}) are 
controlled by the static nonlinearity $u$ with upper critical dimension 
$d_c' = 3$.
Dynamic RG methods had earlier been employed to determine the associated 
critical exponents to two-loop order in a dimensional $\varepsilon = 3 - d$ 
expansion \cite{schmittmann1991,schmittmann1993,praestgaard2000}; the resulting
explicit values are also listed in Table~\ref{ising_table}.
However, numerical data for various finite-size scaling functions by Caracciolo
{\em et al.} favor the JSLC rather than RDLG description \cite{caracciolo2005}.

\begin{figure}[!t]
\includegraphics[angle=0,width=3.4in,clip]{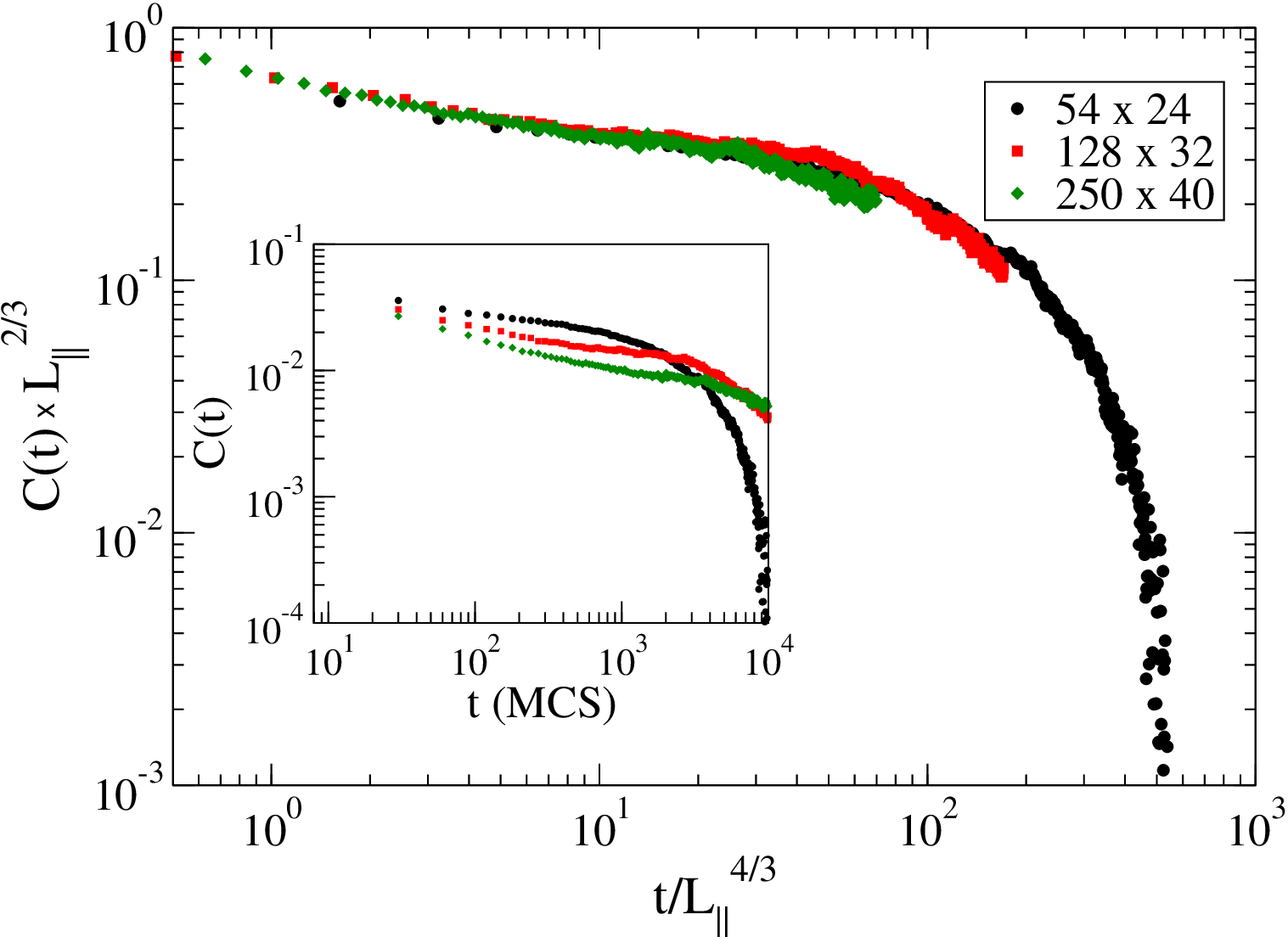}
\caption{({\it color online.}) Finite-size scaling of the steady-state density 
  auto-correlation function for $(L_\parallel,L_\perp) = (54,24)$, $(128,32)$, 
  and $(250,40)$ at the numerically determined ``critical temperatures'' for
  these three system sizes according to Eq.~(\ref{ising_FSS_scaled}); the data 
  are averaged over $2000$, $550$, and $200$ realizations, respectively. 
  The inset depicts the unscaled data.
  \label{crit_fss}}
\end{figure}
We have performed extensive Metropolis Monte Carlo simulations for the critical
KLS model, based on the rates (\ref{hopping_rates}) to numerically determine 
the steady-state scaling exponents.
To account for the anisotropic scaling laws, the simulations should be run with
properly scaled system extensions, where 
$L_\parallel = \mathcal{A} L_\perp^{1+\Lambda}$, where $\Lambda = \Delta$ 
coincides with the correct anisotropy exponent, such that 
$\mathcal{A} = {\rm const}$.
If $\Lambda \neq \Delta$ is chosen, $\mathcal{A}$ does not remain constant and 
will enter the finite-size scaling functions as an additional relevant variable 
\cite{leung1991,caracciolo2005}, see also Ref.~\cite{caracciolo2003b}.
Setting $b = L_\parallel^{-1/(1+\Delta)}$, $\tau = 0$, and letting 
$\vec{x}_\perp,x_{\parallel} \to 0$, Eq.~(\ref{ising_full_scaling_real}) 
reduces to the steady-state finite-size scaling form for the density 
auto-correlation function at criticality,
\begin{equation}
\label{ising_FSS_scaled}
  C(t,L_\parallel) = L_\parallel^{-(d+\Delta-2+\eta)/(1+\Delta)} \, 
  \hat{C}_{\rm FS}\Bigl( t / L_\parallel^{z_\parallel} \Bigr) \, .
\end{equation}
Following Ref.~\cite{wang1996}, we determined the finite-size ``critical 
temperatures'' $T_{\rm C}^{L_\parallel \times L_\perp}$ for three different 
system extensions $(L_\parallel,L_\perp) = (54,24)$, $(128,32)$, and $(250,40)$
(i.e., $\mathcal{A} = 256$ with $\Lambda = 2$) by locating the maximum of the 
variance of the order parameter fluctuations in the steady state, resulting in 
$T_{\rm C}^{54 \times 24} = 0.773$, $T_{\rm C}^{128 \times 32} = 0.782$ and 
$T_{\rm C}^{250 \times 40} = 0.788$ \cite{daquila_thesis_2011}.
As shown in Fig.~\ref{crit_fss}, we thus achieve reasonable but certainly not
unambiguous data collapse using Eq.~(\ref{ising_FSS_scaled}) with the JSLC 
exponents $z_\parallel = \frac43$ and 
$z_\parallel \, \zeta = (d + \Delta - 2 + \eta) / (1 + \Delta) = \frac23$ at 
$d = 2$ (Table~\ref{ising_table}).
We also performed simulations at the critical temperature in the thermodynamic 
limit $T_{\rm C}^\infty = 1.41 \, T_{\rm C}^{\rm eq} = 0.800$ \cite{katz1983,valles1987,leung1991,wang1996,achahbar2001,albano2002,caracciolo2004_a}, 
where $T_{\rm C}^{\rm eq} = 0.5673 \, J$ is the equilibrium critical point of the
square Ising lattice \cite{onsager1944}, and found no noticeable quality 
difference in the scaling collapse.
The data in Fig.~\ref{crit_fss} highlight the difficulty in obtaining accurate 
critical scaling in the steady state; the density correlations display an 
extended crossover regime before reaching their asymptotic behavior 
\cite{caracciolo2004_a}.

\begin{figure}[!t]
  \includegraphics[angle=0,width=3.3in,clip]{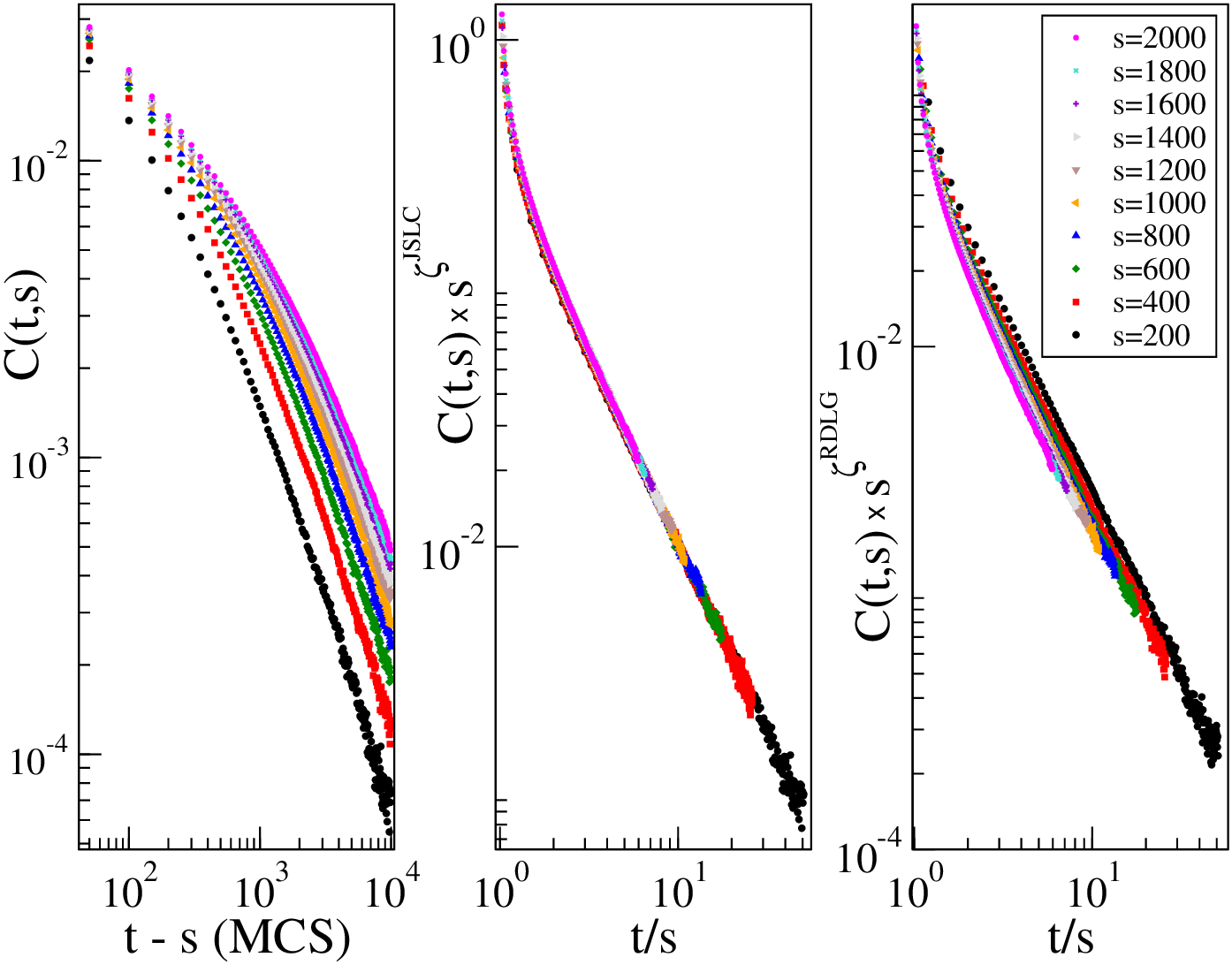}
  \caption{({\it color online.}) Aging scaling plots for the two-time density
  auto-correlation function following a quench to the critical point from a
  high-temperature initial state, on a lattice with dimensions 
  $L_\parallel = 125000$, $L_\perp = 50$ ($\Lambda = \Delta^{\rm JSLC} = 2$).  
  Left panel: unscaled data vs. $t-s$; middle and right panels: scaling plots, 
  Eq.~(\ref{ising_aging_scaling}), using the JSLC and RDLG exponents, 
  respectively.
  Each curve is averaged over $200$ realizations. \label{a_2_age}}
\end{figure}

We therefore next venture to determine the critical exponents via a careful
aging scaling analysis of the density auto-correlation relaxation to the 
non-equilibrium steady state.
Initiating the simulation runs with a random particle distribution allows us to
investigate the out-of-equilibrium relaxation regime wherein time translation
invariance is broken, and two-point correlations become explicit functions of 
two time variables $s$ (referred to as waiting time) and $t > s$
\cite{pleimbook}.
If a system is quenched to the critical point, the initial time sheet may in
principle induce novel infrared singularities, as is the case for the 
relaxational model A with non-conserved order parameter \cite{janssen1989}.
In contrast, for model B and indeed any system with a diffusive conserved order 
parameter field, no new divergences emerge, and the critical initial-slip and 
aging regimes are governed by the steady-state scaling exponents 
\cite{janssen1989,zheng1998,daquila2011}. 
We may therefore simply add the waiting time $s$ as independent scaling 
variable to Eq.~(\ref{ising_full_scaling_real}); setting $b = s^{-1/z}$ then 
yields the simple aging scaling form of the critical density auto-correlation 
function,
\begin{equation}
  C(t,s) = s^{-\zeta} \, f(t/s) \, , \quad 
  \zeta = (d + \Delta - 2 + \eta) / z \ .
\label{ising_aging_scaling}
\end{equation}
Our simulation data for the two-time density auto-correlations at 
$T_{\rm C}^\infty$ are displayed in Fig.~\ref{a_2_age}, for anisotropic lattices
with $\Lambda = \Delta^{\rm JSLC}$; simulation runs with 
$\Lambda = \Delta^{\rm RDLG}$ yield virtually identical results 
\cite{daquila_thesis_2011}.
The left panel, with the unscaled data plotted vs. $t-s$, confirms that time
translation invariance is indeed broken; in the middle and right-hand panel, 
respectively, we attempt to scale the data either using the JSLC or RDLG 
exponent values from Table~\ref{ising_table}.  
The JSLC exponents manifestly yield excellent data collapse, far superior to 
the RDLG values.  
Moreover, in order to determine the aging exponent in an unbiased manner, we
assumed simple aging and obtained $\zeta$ by means of the minimization 
technique described in Ref.~\cite{bhattacharjee2001}, resulting in 
$\zeta \approx 0.48$, in good agreement with the JSLC prediction at $d = 2$.

\begin{figure}[!t]
\includegraphics[angle=0,width=3.4in,clip]{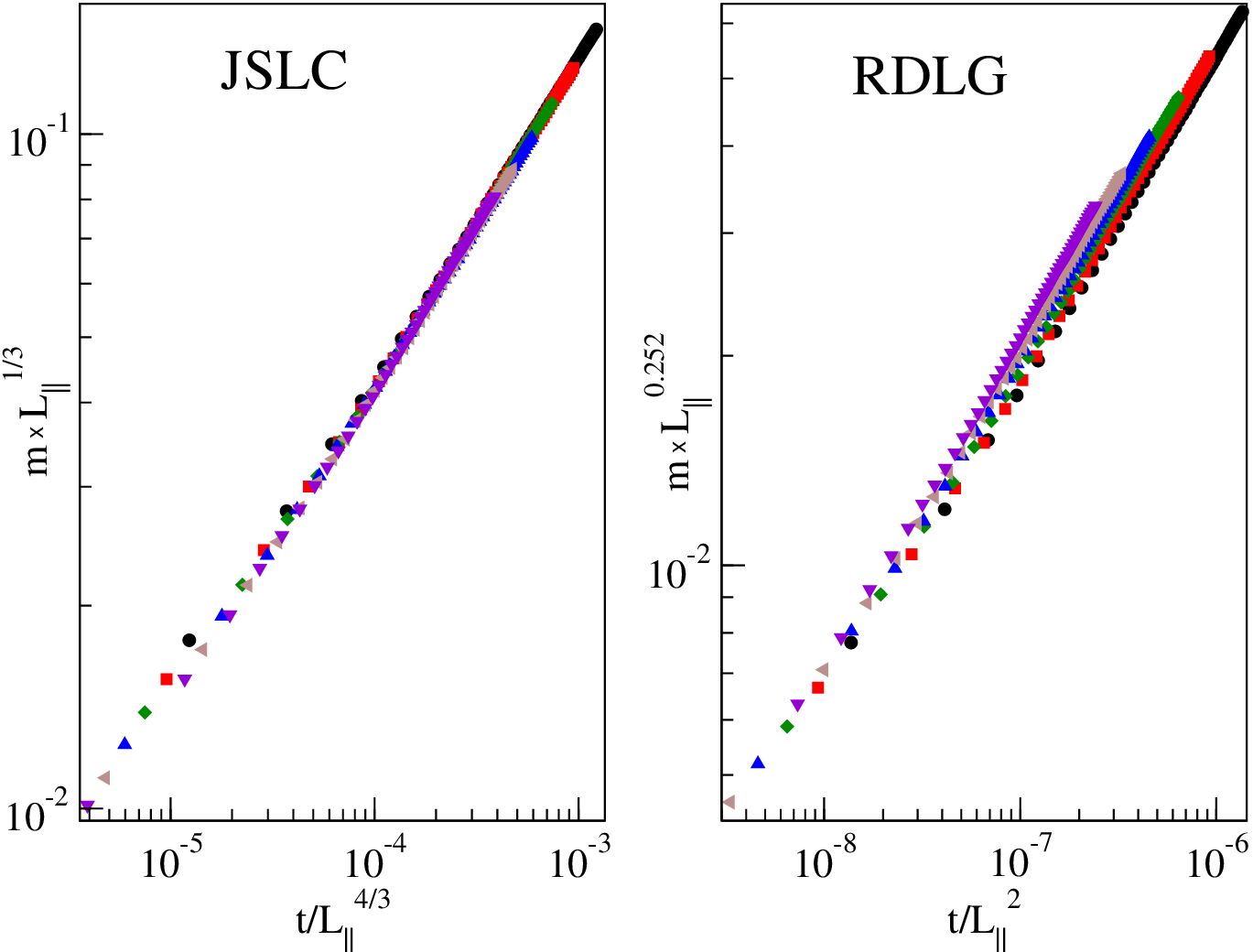}
\caption{({\it color online.}) Finite-size scaling for the order parameter 
  according to Eq.~(\ref{order_fss}).  
  The data and $L_\parallel \times L_\perp$ symbols are the same as in 
  Fig.~\ref{lucky_scale}, which depicts the asymptotic temporal power law.  
  Left panel: excellent data collapse using the JSLC exponents;  
  right panel: attempted data collapse with RDLG values.  
  Each curve is averaged over $500$ realizations. \label{bothscale}}
\end{figure}
Another commonly studied quantity is the anisotropic order parameter, defined 
in the spin representation as
\begin{equation}
\label{order_p_defn}
  m(t,\tau,L_\perp,L_\parallel) = \frac{\sin(\pi/L_\perp)}{2L_{\parallel}} \,
  \bigg| \displaystyle\sum_{j,k=1}^{L_\perp,L_\parallel} \sigma_{j,k}(t) \, 
  e^{\imath \, 2 \pi j / L_\perp} \bigg| \, ,
\end{equation}
which is sensitive to the density modulations transverse to the drive 
\cite{leung1991,wang1996}.
Its finite-size scaling over a range of temperatures near the critical point in
the steady state is known to be sensitive to the precise location of the 
critical temperature and the ``critical region'' chosen for the data analysis 
\cite{leung1996}; stationary finite-size scaling attempts have consequently 
resulted in varying values of the order parameter exponent $\beta$ 
\cite{valles1987,marro1987b,wang1989,leung1991,achahbar1995,marro1996,wang1996,achahbar2001,caracciolo2003,caracciolo2004_a}.
We focus on the temporal evolution of the order parameter (\ref{order_p_defn});
near $T_c$ and beyond mere microscopic time scales, it is governed by the 
universal scaling law \cite{janssen1989,caracciolo2004b}
\begin{equation}
\label{OP_scale_inv}
  m(t,\tau,L_\parallel) = b^{-\beta / \nu_\parallel}\, \breve{m}\Bigl(\!
  b^{-z_\parallel} t, b^{1/\nu_\parallel} \tau, b^{1/(1+\Delta)} L_\perp^{-1},
  b L_\parallel^{-1}\!\Bigr) .
\end{equation}
Identifying $b = L_\parallel$ yields the critical finite-size scaling
\begin{equation}
\label{order_fss}
  m(t,L_\parallel) = L_\parallel^{-\beta / \nu_\parallel} \, 
  \hat{m}\Bigl( t/L_\parallel^{z_\parallel}\Bigr) \ .
\end{equation}
We present our Monte Carlo simulation results in Fig.~\ref{bothscale}, 
employing two-dimensional anisotropic lattices with 
$\Lambda = \Delta^{\rm JSLC}$.
We attempted to scale the data via Eq.~(\ref{order_fss}) using both JSLC and 
RDLG exponent values, see Table~\ref{ising_table}; the JSLC critical exponents 
$z_\parallel = \frac43$ and $\beta / \nu_\parallel = \frac13$ clearly give 
superior data collapse in this ``initial slip'' region.

\begin{figure}[!t]
\includegraphics[angle=0,width=3.4in,clip]{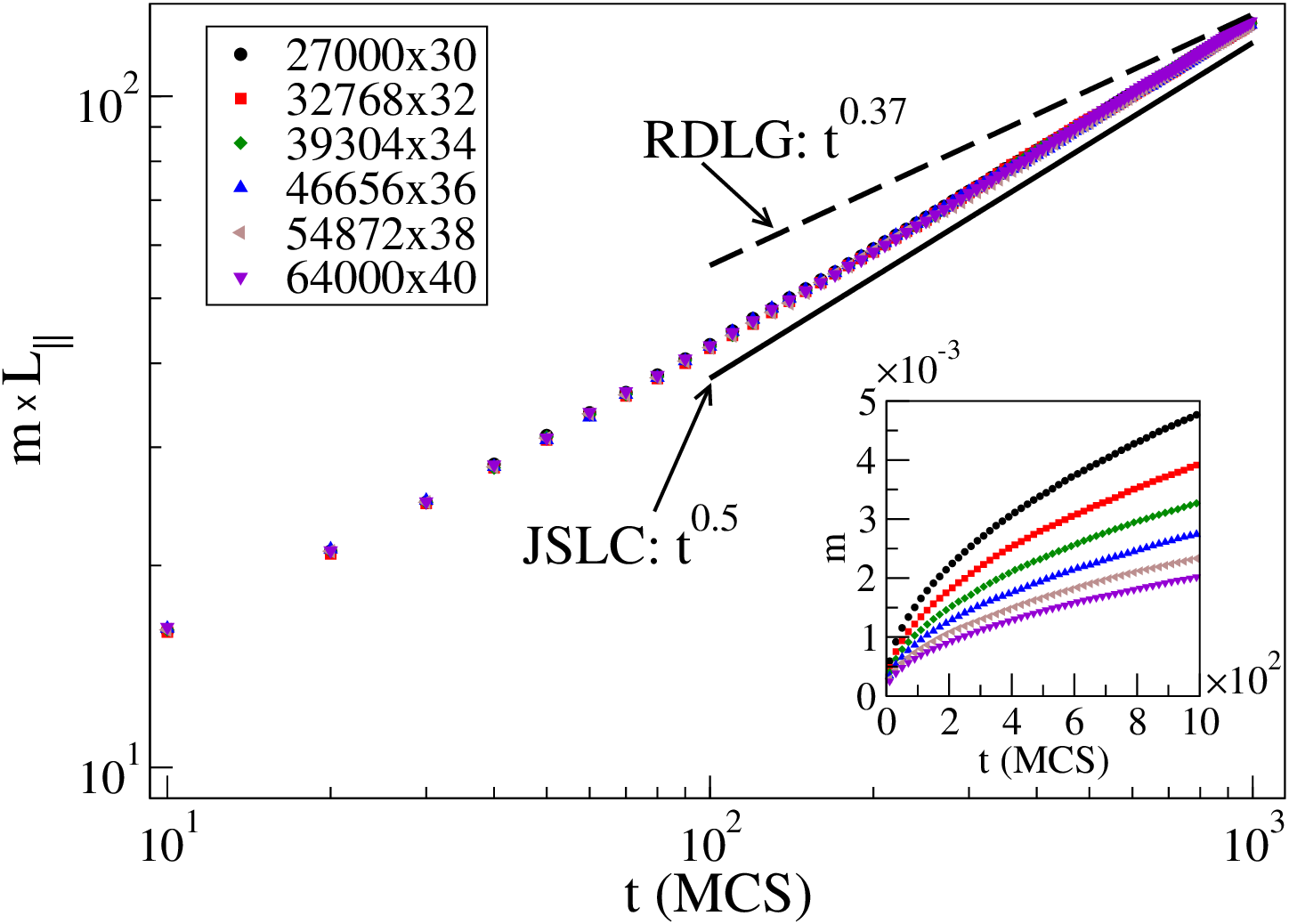}
\caption{({\it color online.}) Scaling plot according to 
  Eq.~(\ref{OP_scale_linear}).  
  The dashed / solid straight lines above / below the simulation data 
  respectively correspond to the predicted RDLG / JSLC power laws.
  The inset shows the unscaled order parameter growth.  
  Each curve is averaged over $500$ realizations. \label{lucky_scale}}
\end{figure}

To extract the asymptotic temporal power law in the initial slip regime, which
allows us to investigate very large systems, we set $b = t^{1/z_\parallel}$ and 
$\tau = 0$ in Eq.~(\ref{OP_scale_inv}), whence 
$m(t,L_\parallel) = t^{-\beta / z_\parallel \nu_\parallel} \, \widetilde{m}
\Bigl( t^{1/z} L_\perp^{-1}, t^{1/z_\parallel} L_\parallel^{-1} \Bigr)$.
Substituting $L_\perp = \mathcal{A} \, L_\parallel^{1/(1+\Delta)}$ with 
$\mathcal{A} = {\rm const}$, $\widetilde{m}$ becomes a function of 
$x = t^{1/z_\parallel} L_\parallel^{-1}$ only.  
Following the procedure outlined in Ref.~\cite{zheng1998}, we expand the 
regular scaling function 
$\widetilde{m}(x) = \widetilde{m}_0 + \widetilde{m}_1 x + \ldots$ for 
$x \ll 1$; here $\tilde{m}_0 = 0$ as alternating occupied/empty initial 
conditions were chosen.
Ignoring higher-order terms, we arrive at
\begin{equation}
\label{OP_scale_linear}
  m(t,L_\parallel) \, L_\parallel \sim t^{\kappa} \, , \quad 
  \kappa = (1 - \beta / \nu_\parallel) / z_\parallel \ .
\end{equation}
Figure~\ref{lucky_scale} demonstrates convincing data collapse with this 
scaling form.  
Performing a fit to the $64000 \times 40$ lattice data yields 
$\kappa \approx 0.487$, close to the JSLC value $\frac12$ \cite{gustavo2003}.

In summary, we present detailed Monte Carlo simulation results for the KLS 
model at criticality confirming the JSLC scaling exponents through the 
measurement of various dynamical quantities, specifically the order parameter 
in the initial slip and the two-time density auto-correlation function in the
aging regime, which can be numerically accessed for very large lattices.
Finite-size scaling of the density auto-correlation function gives satisfactory
(if not fully convincing) data collapse with the JSLC exponents.  
Simple scaling arguments allow us to relate the aging exponent to known 
steady-state critical exponents \cite{janssen1989,caracciolo2004b}.  
We have performed Monte Carlo simulations in the transient regime where time 
translation invariance is broken, and measured the density auto-correlation at 
different waiting times, thus exploring in detail the universal relaxation 
features towards a critical non-equilibrium stationary state.  
The aging exponent inferred from the JSLC exponent values yields convincing 
data collapse over a range of waiting times, even with different lattice 
anisotropies.  
In contrast, the RDLG scaling exponents do not permit a reliable simple aging 
scaling collapse.  
The time-dependent order parameter data at short times for different lattice 
sizes were found to fit a universal master curve using finite-size scaling with
JSLC exponents.  
Applying the same scaling form with RDLG exponents again could not collapse our
data.  
Supporting the conclusions from a careful numerical finite-size scaling 
analysis \cite{caracciolo2005}, our work thus provides strong evidence that the 
universal features of the critical KLS model are (even with infinite drive) 
adequately described by the JSLC coarse-grained stochastic Langevin equation, 
and captured by the associated field-theoretic RG approach.

\begin{acknowledgments}
This material is based upon work supported by the U.S. Department of Energy,
Office of Basic Energy Sciences under Award Number DE-FG02-09ER46613. 
We thank Ulrich Dobramysl, Thierry Platini, Michel Pleimling, Beate 
Schmittmann, and Royce Zia for helpful discussions and insightful suggestions.
\end{acknowledgments}

\bibliographystyle{h-physrev}

\end{document}